\documentclass{Interspeech}

\interspeechcameraready

\title{Beyond Manual Transcripts: The Potential of Automated Speech Recognition Errors in Improving Alzheimer’s Disease Detection\thanks{* Corresponding author.}}
\author[affiliation={1}]{Yin-Long}{Liu}
\author[affiliation={1}]{Rui}{Feng}
\author[affiliation={1}]{Jia-Xin}{Chen}
\author[affiliation={1}]{Yi-Ming}{Wang}
\author[affiliation={1,2*}]{Jia-Hong}{Yuan}
\author[affiliation={1,2}]{Zhen-Hua}{Ling}

\affiliation{National Engineering Research Center of Speech and Language Information Processing}{University of Science and Technology of China}{Hefei, P. R. China}
\affiliation{Interdisciplinary Research Center for Linguistic Sciences}{University of Science and Technology of China}{Hefei, P. R. China}

\email{lyl2001@mail.ustc.edu.cn, \{jiahongyuan, zhling\}@ustc.edu.cn}

\usepackage{comment}
\usepackage{cite}
\usepackage{hyperref}
\usepackage{algorithmic}
\usepackage{textcomp}
\usepackage{xcolor}
\usepackage{enumerate}
\usepackage{cleveref}
\usepackage{graphicx}    % 处理图像
\usepackage{subcaption}  % 用于子图
\usepackage{makecell}
\usepackage{enumitem}
\usepackage[bottom]{footmisc} 
\usepackage{bm}
\usepackage{adjustbox}
\usepackage{amsmath,amsthm,amssymb,amsfonts}
\usepackage{bigstrut,multirow,rotating,booktabs}%表格插件 
\begin{document}

\maketitle
\begin{abstract}
Recent breakthroughs in Automatic Speech Recognition (ASR) have enabled fully automated Alzheimer’s Disease (AD) detection using ASR transcripts. Nonetheless, the impact of ASR errors on AD detection remains poorly understood. This paper fills the gap. We conduct a comprehensive study on AD detection using transcripts from various ASR models and their synthesized speech on the ADReSS dataset. Experimental results reveal that certain ASR transcripts (ASR-synthesized speech) outperform manual transcripts (manual-synthesized speech) in detection accuracy, suggesting that ASR errors may provide valuable cues for improving AD detection. Additionally, we propose a cross-attention-based interpretability model that not only identifies these cues but also achieves superior or comparable performance to the baseline. Furthermore, we utilize this model to unveil AD-related patterns within pre-trained embeddings. Our study offers novel insights into the potential of ASR models for AD detection.
\end{abstract}
\vspace{-0.3cm}
\section{Introduction}
\vspace{-0.2cm}
\label{sec:Introduction}
%介绍AD, text和audio进行AD检测, ASR trans进行AD检测引出没有一个研究来进行全面的ASR trans研究, SSL黑盒特性进而可解释性, 本文的工作
Alzheimer’s Disease (AD), the leading cause of dementia, is a progressive neurodegenerative disorder causing irreversible brain damage and cognitive decline in memory, language, attention, and executive function \cite{nestor2004advances}. Early detection is crucial, and compared to traditional clinical methods, speech-and-language-based automatic AD diagnosis has gained increasing attention for its non-invasive, cost-effective, and convenient nature \cite{talkar24_interspeech,liu24f_interspeech,botelho24_interspeech,pan2021using,mei2023ustc,liu2024leveraging}, further advanced by recent challenges \cite{ luz2020fuente,luz2021detecting,luz2023multilingual,garcia2024connected}.

Studies show that linguistic features often outperform acoustic features alone in AD detection \cite{syed2020automated,li2021comparative}. However, manual transcripts is labor-intensive, making it impractical for large-scale datasets. In contrast, Automatic Speech Recognition (ASR) offers efficient, rapid transcripts, reducing manual effort. Recent ASR advancements, such as Wav2Vec2 \cite{baevski2020wav2vec}, HuBERT \cite{hsu2021hubert}, WavLM \cite{chen2022wavlm}, and Whisper \cite{radford2023robust}, have achieved remarkable performance, facilitating fully automated AD detection using ASR transcripts. Although previous studies suggest ASR transcripts may underperform compared to manual ones \cite{zhou2016speech,balagopalan2019impact}, a comprehensive evaluation of various ASR models for AD detection is limited \cite{li2024useful,liu2025can}. This study addresses that gap by fine-tuning 18 variants of the four ASR models on three datasets from DementiaBank (DB) and evaluating them on the ADReSS challenge dataset \cite{luz2020fuente}, generating 36 ASR transcripts with varying Word Error Rate (WER) for each sample (18 from the original models and 18 from the fine-tuned ones).

Pre-trained language models like BERT \cite{devlin2018bert} can capture complex patterns for detecting cognitive decline and have proven effective in AD detection \cite{ wang22l_interspeech, yuan2020disfluencies}. We extract BERT embeddings from ASR and manual transcripts of the ADReSS dataset and design a self-attention-based model for AD detection. Our results show that certain ASR transcripts outperform manual ones, suggesting that ASR errors may offer valuable cues for distinguishing AD from Healthy Controls (HC). 
To further explore this, we synthesize speech from ASR and manual transcripts using the cutting-edge Text-to-Speech (TTS) model CosyVoice2 \cite{du2024cosyvoice}. Based on the effectiveness of the pre-trained speech model Wav2Vec2 \cite{baevski2020wav2vec} in dementia detection \cite{li2023leveraging,pan2025two}, we extract Wav2Vec2 embeddings from these synthesized speech and perform AD detection again. Remarkably, some ASR-synthesized speech, exhibits higher AD detection performance than the manual-synthesized speech. To our knowledge, no previous work has reported or explained this phenomenon, making it worthy of an in-depth investigation. 
 % (speech synthesized from ASR transcripts) (speech synthesized from manual transcripts)

\begin{figure*}[t!]
    \centering
\includegraphics[width=1\linewidth]{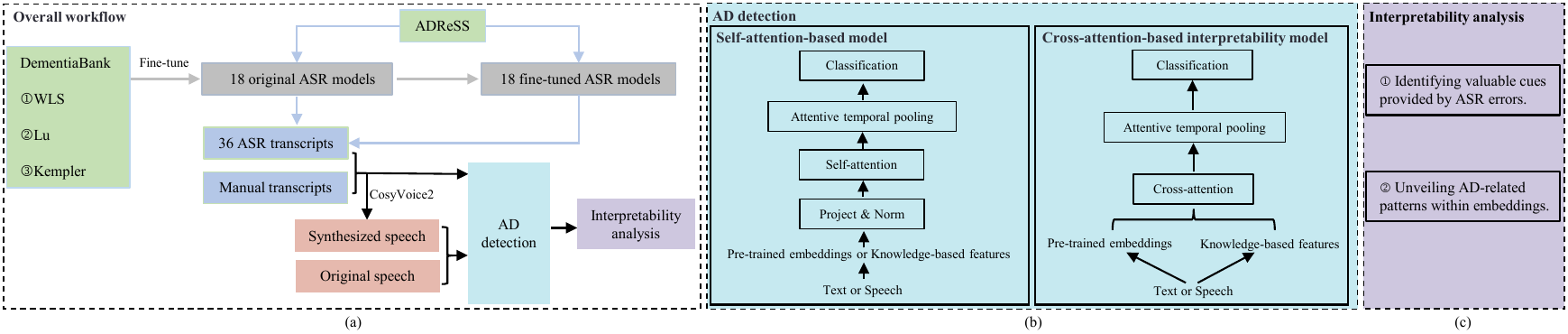}
\vspace{-0.6cm}
    \caption{(a) refers to the overall workflow, (b) refers to the AD detection models, and (c) refers to the interpretability analysis.}
    \label{fig:Over workflow.}
     \vspace{-0.7cm}
\end{figure*}

Understanding the decision-making process of deep learning-based medical systems is crucial for fostering transparency and trust in clinical applications. Iqbal et al. \cite{iqbal2024explainable} used interpretability tools to identify key linguistic features in AD detection. Gimeno-Gómez et al. \cite{gimeno2024unveiling} developed a framework to uncover meaningful speech characteristics for Parkinson's disease detection. Inspired by these works, we propose a cross-attention-based interpretability model to explore why certain ASR transcripts (or synthesized speech) outperform manual ones in AD detection accuracy, and to elucidate the underlying mechanisms driving classification.
% , and to elucidate the underlying mechanisms driving classification. 
Specifically, we first select some knowledge-based text and speech features (also used individually for AD detection), which are then fused with embeddings from BERT or Wav2Vec2 within our model. By analyzing attention scores from the best-performing models based on ASR and manual transcripts (or ASR- and manual-synthesized speech), we identify which knowledge-based features are enhanced by ASR errors, leading to their recognition as valuable cues for AD detection, ultimately improving classification accuracy. We also employ this model to unveil AD-related patterns withih pre-trained embeddings. The overall workflow is illustrated in Figure~\ref{fig:Over workflow.}. Our main contributions are as follows:

% Our main contributions can be summarized as follows:
\begin{itemize}
\vspace{-0.1cm}
\item 
% This study is the first comprehensive investigation of AD detection using ASR transcripts with varying WER and their synthesized speech.
As the first comprehensive study to investigate AD detection using ASR transcripts with varying WER and their synthesized speech, advancing ASR-assisted AD detection.
\item 
We find that certain ASR transcripts and ASR-synthesized speech outperform their manual counterparts, suggesting that ASR errors may provide valuable cues for AD detection.
\item 
We propose a cross-attention-based interpretability model that identifies valuable cues from both transcript and speech perspectives, achieving superior or comparable performance and unveiling AD-related patterns within embeddings.
\end{itemize}
% \begin{table}[b]
% \vspace{-0.6cm}
%   \caption{Statistical details of the datasets.}
%     % \textbf{Different subdataset}
%     \label{tab:Statistical Information}
% \centering
% \resizebox{1.0\linewidth}{!}{
% \begin{tabular}{c cc cc}
% \toprule
% \multirow{2}{*}{Statistical Values} & \multicolumn{2}{c}{ASR} & \multicolumn{2}{c}{ADReSS} \\
% \cmidrule(lr){2-3} \cmidrule(lr){4-5}
% & \multicolumn{1}{c}{Training Set} & \multicolumn{1}{c}{Test Set} & \multicolumn{1}{c}{Training Set} & \multicolumn{1}{c}{Test Set} \\

% \midrule
% Num of subject-level samples    & 196                    & 49                       & 108                      & 48                       \\

% Num of utterance-level segments & 3100                   & 847                      & 1478                     & 584                      \\

% Mean / total duration of utterance-level segments    & 3.4s / 2.93h & 3.22s / 0.76h & 4.22s / 1.73h & 5.43s / 0.88h \\
% \bottomrule
% \end{tabular}}
% \end{table}
\vspace{-0.4cm}
\section{Dataset}
\vspace{-0.3cm}
%介绍ADReSS数据集和微调ASR数据集以及合成的语音数据？，参考ICASSP
We selected and organized three datasets from DB for fine-tuning the ASR models: WLS (187 samples), Lu (54 samples), and Kempler (4 samples). The ADReSS challenge dataset \cite{luz2020fuente} from Interspeech 2020 was used for binary AD detection, with 108 training samples (54 AD, 54 HC) and 48 test samples (24 AD, 24 HC). Each sample consists of an audio recording and a corresponding manual transcript, which include subject's verbal descriptions of the ``Cookie Theft" picture elicited by the interviewer. The manual transcripts were annotated in the CHAT format, which we processed to align with the subjects’ actual speech. The WLS, Lu, and Kempler datasets were merged, resulting in 245 samples, which were split into ASR training (80\%) and test sets (20\%). Using timestamps from the CHAT annotations, we segmented the audio recordings into utterances that contained only the subjects' speech.
% The statistical details of these datasets are summarized in Table~\ref{tab:Statistical Information}.
We fine-tuned the ASR models on these utterance-level segments and transcripts. For ADReSS, ASR subject-level transcripts were created by concatenating the ASR utterance-level transcripts. The ASR subject-level transcripts (synthesized speech), manual transcripts (synthesized speech), and original speech of ADReSS were used for subsequent AD detection.
\vspace{-0.3cm}
\section{Methods}
\vspace{-0.2cm}
\subsection{Fine-tuning ASR models}
\vspace{-0.2cm}
We selected 18 variants from Wav2Vec2 \cite{baevski2020wav2vec}, HuBERT \cite{hsu2021hubert}, WavLM \cite{chen2022wavlm}, and Whisper \cite{radford2023robust} for fine-tuning, as these models provide varying WER. The selected variants are:
\textit{\textbf{wav2vec2}-\{base-100h, base-960h, large-960h, large-960h-lv60, large-960h-lv60-self, large-xlsr-53-english, xls-r-1b-english}\},  \textit{\textbf{hubert}-\{large-ls960-ft, xlarge-ls960-ft}\}, \textit{\textbf{wavlm}-libri-clean-100h-\{base-plus, large}\}, \textit{\textbf{whisper}-\{tiny, base, small, medium, large, large-v2, large-v3}\}.
% \begin{itemize}
% \vspace{-0.1cm}
% \item{\textit{wav2vec2-\{base-100h, base-960h, large-960h, large-960h-lv60, large-960h-lv60-self, large-xlsr-53-english, xls-r-1b-english}}\}
% \item{\textit{hubert-\{large-ls960-ft, xlarge-ls960-ft}}\}
% \item{\textit{wavlm-libri-clean-100h-\{base-plus, large}}\}
% \item{{\small\textit{whisper-\{tiny, base, small, medium, large, large-v2, large-v3}}\}}
% \vspace{-0.1cm}
% \end{itemize}
All these ASR models are available on HuggingFace, and we fine-tuned them using the standard methods from the \textit{Transformers} Python library.
% For fine-tuning Wav2Vec2, HuBERT, and WavLM, we employed the \textit{Wav2Vec2Processor} and \textit{AutoModelForCTC} packages. For fine-tuning Whisper, we utilized the \textit{WhisperForConditionalGeneration} and \textit{WhisperProcessor} packages. All these packages are sourced from the \textit{Transformers} library\footnote{https://github.com/huggingface/transformers}.
\vspace{-0.3cm}
\subsection{Speech synthesis}
\vspace{-0.2cm}
To explore the impact of ASR transcripts with varying WER on AD detection from a speech perspective, we utilized CosyVoice2 \cite{du2024cosyvoice} to synthesize speech from both ASR and manual transcripts. CosyVoice2 is an advanced TTS model that delivers nearly lossless synthesis. Using its zero-shot functionality, we randomly selected a segment from the subject’s original speech as a prompt for each transcript, ensuring the synthesized speech closely matched the original characteristics.
% To investigate the impact of ASR transcripts with varying WER on AD detection from the perspective of speech, we employed CosyVoice2 \cite{du2024cosyvoice} to synthesize speech from both ASR and manual transcripts. CosyVoice2 is an advanced model for scalable streaming speech synthesis, achieving low-latency, high naturalness, and near-lossless speech synthesis through enhanced quantization techniques and the use of pre-trained large language models. Following the zero-shot usage\footnote{https://github.com/FunAudioLLM/CosyVoice} of CosyVoice2, for each transcript, the corresponding prompt speech is required to ensure that the synthesized speech mimics the characteristics contained in the prompt speech. In this context, for each transcript, we randomly selected a segment from the original speech of the corresponding subject as the prompt speech, ensuring that the characteristics of the synthesized speech align with those of the subject's original speech.

\vspace{-0.3cm}
\subsection{Feature extraction}
\label{sec:Feature extraction}
\vspace{-0.2cm}
\subsubsection{Knowledge-based features}
\vspace{-0.2cm}
\label{sec:Knowledge-based features}
To enable interpretable analysis, we extracted 35 and 60 knowledge-based features from each text and speech sample, respectively, chosen for their proven effectiveness as AD biomarkers in prior studies \cite{ iqbal2024explainable,gimeno2024unveiling,pereztoro21_interspeech}. 
% The 35 knowledge-based text features were classified into following six categories:
% The 35 knowledge-based text features, extracted using the lexical-diversity, textstat, spaCy libraries, and the MRC psycholinguistic database, are classified into six categories:

The 35 knowledge-based text features, extracted using the Python libraries \textit{lexical-diversity}, \textit{textstat}, and \textit{spaCy}, as well as the MRC psycholinguistic database \cite{wilson1988mrc}, were classified into six categories: \textbf{Lexical Diversity} (4): type-token ratio (TTR), mean segmental TTR (MSTTR), moving average TTR (MATTR), measure of lexical textual diversity (MTLD); \textbf{Content Complexity} (7): syllable count, stop words count, lexicon count, difficult words count, average sentence length, content density, propositional density; \textbf{Disfluencies} (2): the ratio of filler pauses (uh, um, er, and ah) and repetition words; \textbf{Part of Speech} (11): the ratio of Pronouns, Verbs, Nouns, Adjectives, Adverbs, Conjunctions, Articles, Determiners, Prepositions, Pronoun-Verb, and Pronoun-Noun; \textbf{Readability} (6): Flesch Reading Ease (FRE), Flesch-Kincaid Grade Level (FKGL), Gunning Fog Index (GFI), Coleman-Liau Index (CLI), Dale-Chall Readability Score (DCRS), Automated Readability Index (ARI); \textbf{Psycholinguistic Features} (5): familiarity, concreteness, imagability, meaningfulness, age of acquisition rating.

The 60 knowledge-based speech features, extracted using the DisVoice toolkit\footnote{https://github.com/jcvasquezc/DisVoice}, were classified into four categories (the full description is in the toolkit above): \textbf{Articulation} (4): the mean and standard deviation (std) of the first (F1) and second (F2) formant frequencies; \textbf{Glottal} (14): the mean and std of seven descriptors: variability of time between consecutive glottal closure instants (GCI), average and variability of normalized amplitude quotient (NAQ) for consecutive glottal cycle, average and variability of opening quotient (OQ) for consecutive glottal cycles, average and variability of harmonic richness factor (HRF); \textbf{Phonation} (7): the mean of jitter, shimmer, amplitude perturbation quotient (APQ), pitch perturbation quotient (PPQ), and logarithmic energy (logE); the mean and std of the first derivative of the fundamental frequency (DF0); \textbf{Prosody} (35): the mean, std, maximum, minimum, skewness, and kurtosis of F0 contour; the mean, std, skewness, and kurtosis of energy contour for voice segments; number of voiced segments per second (NVSS);  the mean, std, skewness, kurtosis, maximum, and minimum of voiced duration, unvoiced duration, and pause duration; the six duration ratios: pause/(voiced+unvoiced) (PVU), pause/unvoiced (PU), unvoiced/(voiced+unvoiced) (UVU), voiced/(voiced+unvoiced) (VVU), voiced/pause (VP), unvoiced/pause (UP).

\vspace{-0.3cm}
\subsubsection{Pre-trained embeddings}
\vspace{-0.2cm}
\label{sec:Pre-trained embeddings}
To capture rich characteristics for AD detection, we employed pre-trained BERT\footnote{google-bert/bert-large-uncased} and Wav2Vec2\footnote{jonatasgrosman/wav2vec2-large-xlsr-53-english} models from HuggingFace to extract embeddings. Both models consisted of 24 transformer layers. Previous studies \cite{li2023leveraging, gimeno2024unveiling, jawahar2019does, pasad2021layer} have shown that Wav2Vec2's intermediate layers capture more acoustic and phonetic information, while BERT’s intermediate layers contain more syntactic information, offering superior AD detection performance compared to other layers. These findings align with our preliminary experiments. Therefore, we focused on the 8th layer of Wav2Vec2 and the 11th layer of BERT to better fuse these embeddings with knowledge-based features. For each text, we used BERT to generate embeddings of dimension \(N \times 1024\), where \(N\) was the number of tokens (excluding [CLS] and [SEP]). For speech, we segmented each sample into 30-second chunks, extracted features with Wav2Vec2, and concatenated the segment-level embeddings along the time dimension to form subject-level embeddings of dimension \(T \times 1024\), where \(T\) was the number of time steps.
\begin{table}[t!]
% \vspace{-0.5cm}
\caption{Mean WER (\%) of ASR models on ADReSS dataset.}
\label{tab:ASR models performance}
\centering
\resizebox{1.0\linewidth}{!}{
\begin{tabular}{cccc}
\toprule
ASR models      & Original / Fine-tuned & ASR models       & Original / Fine-tuned \\
\cmidrule(lr){1-2} \cmidrule(lr){3-4}
w2v100          & 68.55 / 54.32         & wavlm base       & 68.53 / 59.22         \\
w2v960          & 61.05 / 50.80         & wavlm large      & 57.76 / 54.13         \\
w2v960 large    & 57.16 / 45.99         & whisper tiny     & 62.58 / 43.33         \\
w2v960 large lv & 51.86 / 45.87         & whisper base     & 58.10 / 36.39         \\
w2v960 self     & 49.86 / 42.65         & whisper small    & 49.91 / 30.47         \\
w2v xlsr        & 55.67 / 42.19         & whisper medium   & 45.70 / 29.39         \\
w2v xlsr 1b     & 49.89 / 36.79         & whisper large    & 45.32 / 28.04         \\
hubert large    & 50.23 / 41.89         & whisper large v2 & 45.98 / 28.65         \\
hubert xlarge   & 49.18 / 45.99         & whisper large v3 & 46.00 / \textbf{26.36 }       \\
\bottomrule
\end{tabular}}
\vspace{-0.7cm}
\end{table}
\vspace{-0.3cm}
\subsection{AD detection models}
\vspace{-0.2cm}
The proposed AD detection models are illustrated in Figure~\ref{fig:Over workflow.}.
% In this section, we describe the two proposed AD detection models, whose architectures are shown in Figure~\ref{fig:Over workflow.}.
\vspace{-0.3cm}
\subsubsection{Self-attention-based model}
\vspace{-0.2cm}
The model takes pre-trained embeddings or knowledge-based features as input. It includes a linear projection with layer normalization, a self-attention module, an attentive temporal pooling module \cite{li2023leveraging, santos2016attentive} for capturing richer temporal feature statistics, and a linear classification module. This model serves as a baseline for comparison with the cross-attention-based model.
\vspace{-0.3cm}
\subsubsection{Cross-attention-based interpretability model}
\vspace{-0.2cm}
\label{sec:Cross-attention-based interpretability model}
The model utilizes knowledge-based features and pre-trained embeddings as inputs. According to Section~\ref{sec:Feature extraction}, for each sample, we represented the knowledge-based feature as \( X_{\text{kno}} \in \mathbb{R}^{1 \times m} \) and the pre-trained embedding as \( X_{\text{emb}} \in \mathbb{R}^{n \times 1024} \), where \( m \) was either 35 or 60, and \( n \) corresponded to the number of tokens or time steps. We modified the standard cross-attention mechanism to effectively integrate both representations while providing interpretability, as shown in Equation~\ref{equation:cross attention}. 
Specifically, we first repeated \( X_{\text{kno}} \) \( n \) times to obtain \( X_{\text{kno}}^{\prime} \in \mathbb{R}^{n \times m} \). Then, we defined \( Q = X_{\text{emb}}\cdot W_q \), \( K = X_{\text{kno}}^{\prime} \cdot W_k \), and \( V = X_{\text{emb}} \cdot W_v \), where \( W_q \), \( W_v \in \mathbb{R}^{1024 \times 1024} \) were learnable weight matrices, and \( W_k \) was an identity matrix \( I \). As a result, the knowledge-based features directly served as \(K\), providing a set of interpretable dimensions that were fused with the embeddings. \( d_k \) was the dimensionality of \( K \), which was \( m \). Next, the attention score matrix \( A \in \mathbb{R}^{1024 \times m} \) was computed via the scaled dot-product attention and softmax, capturing the relationship between the pre-trained embeddings and knowledge-based features. We utilized this matrix for subsequent interpretability analysis. 
Finally, the attention scores were applied to \( V \), resulting in an enriched representation \( Y \in \mathbb{R}^{n \times m} \), which was passed through the attentive temporal pooling module and a linear classifier to produce the final prediction.
\vspace{-0.4cm}
\begin{align}
  Attention(Q, K, V) &= V \cdot softmax\left(\frac{Q^T K}{\sqrt{d_k}}\right)
  \label{equation:cross attention}
\end{align}
\vspace{-0.4cm}
\begin{table}[t!]
    \caption{AD detection mean accuracy (\%) on transcripts (manual + 36 ASR) and speech (original + synthesized from manual and 36 ASR transcripts). The three values separated by ``/" represent the accuracy of three AD detection methods: from left to right, they correspond to inputting knowledge-based features / pre-trained embeddings into the self-attention model, and using the cross-attention model. In each AD detection method, the bold accuracy from ASR transcripts (ASR-synthesized speech) indicates its superiority over the corresponding manual transcripts (manual-synthesized speech).}
    %The three values separated by '/' represent accuracy under different conditions: from left to right, they correspond to inputting knowledge-based features and  pre-trained embeddings into the self-attention model ,and using the cross-attention model.
    % \vspace{-0.1cm}
    \label{tab:AD detection results}
    \centering
\resizebox{1.0\linewidth}{!}{
\begin{tabular}{cllll}
\toprule
                                 & \multicolumn{2}{c}{Transcripts}                               & \multicolumn{2}{c}{Speech}     \\   

\cmidrule(lr){2-3} \cmidrule(lr){4-5}
\cmidrule(lr){2-3} \cmidrule(lr){4-5}

                         & \multicolumn{2}{c}{Manual transcripts: 79.17 / 81.67 / 80.42}      & \multicolumn{2}{c}{\begin{tabular}[c]{@{}c@{}}Original Speech: 75.83 / 77.92 / 78.75\\     
                                                                                                                                    Manual-synthesized: 59.58 / 74.58 / 70.42\end{tabular}} \\
\midrule 
\midrule
ASR models                       & \multicolumn{1}{c}{Fine-tuned} & \multicolumn{1}{c}{Original} & \multicolumn{1}{c}{Fine-tuned}                                                    & \multicolumn{1}{c}{Original}                                                   \\
\midrule
w2v100                           & 75.42 / 81.25 / 75.83          & 72.92 / 74.17 / 72.08        & 57.92 / 68.75 / 62.08      & 59.17 / 72.50 / 62.92     \\
w2v960                           & 75.83 / 77.92 / 74.17          & 76.25 / 76.67 / 76.25        & \textbf{65.42} / 69.58 / 60.42       & \textbf{66.67} / 67.08 / 65.83     \\
w2v960 large                     & 72.92 / 79.58 / 73.33          & 75.42 / 77.08 / 75.00        & 57.92 / 67.50 / 63.75        & 55.42 / 68.33 / 63.33    \\
w2v960 large lv                  & 76.67 / 76.67 / 77.50          & 69.58 / 74.17 / 72.08        & 55.42 / 65.83 / 64.17       & 55.00 / 67.08 / 59.17       \\
w2v960 self                      & \textbf{81.25} / 80.83 / 79.58  & 75.83 / 77.92 / 75.42       & \textbf{66.67} / 71.67 / 67.50       & \textbf{69.58} / 70.83 / 64.17    \\
w2v xlsr                         & 74.17 / 75.83 / 73.33          & 72.50 / 77.08 / 69.38        & \textbf{62.08} / 69.17 / 60.83       & 52.92 / 70.42 / 58.96 \\
w2v xlsr 1b                      & 78.33 / 75.42 / 77.08          & 77.92 / 79.58 / 74.58        & \textbf{60.83} / 72.50 / 65.42      & \textbf{62.08} / 68.75 / 62.08       \\
hubert large                     & 71.25 / \textbf{81.67} / 76.67  & \textbf{81.25} / 78.75 / 74.38  & 59.17 / 72.08 / 63.33    & \textbf{63.75} / 72.92 / 57.71                            \\
hubert xlarge                    & 78.75 / 80.42 / 78.33          & 77.50 / 79.17 / 75.42        & \textbf{62.08} / 70.00 / 65.83      & \textbf{63.75} / 65.00 / 58.96           \\
wavlm base                       & 78.75 / 77.50 / 76.25          & 77.50 / 75.83 / 75.00        & \textbf{63.75} / 64.17 / 59.58      & \textbf{63.33} / 69.58 / 57.92               \\
wavlm large                      & 76.25 / \textbf{82.50} / \textbf{81.25}  & 75.42 / 75.42 / 76.25  & \textbf{68.75} / 70.42 / 66.67       & 53.75 / 72.50 / 65.00         \\
whisper tiny                     & 77.50 / 78.75 / 79.17     & \textbf{79.58} / 75.00 / 71.25          & \textbf{62.08} / 72.08 / 67.50   & \textbf{65.42} / 63.33 / 66.67    \\
whisper base                     & \textbf{79.17} / 79.58 / \textbf{80.83}        & 72.50 / 76.67 / 72.92        & \textbf{65.42} / 70.42 / 68.75   & \textbf{67.50} / 67.50 / 64.17    \\
whisper small                    & \textbf{80.00} / \textbf{82.50} / \textbf{81.25}  & \textbf{80.42} / \textbf{82.08} / 76.67  & \textbf{69.58} / \textbf{75.42} / \textbf{72.08}   & \textbf{64.58} / 65.00 / 65.83      \\
whisper medium                   & \textbf{80.42} / \textbf{82.92} / 78.33       & \textbf{85.42} / \textbf{81.67} / 79.58   & \textbf{62.92} / \textbf{75.83} / \textbf{71.67} & \textbf{60.00} / \textbf{75.00} / 67.08    \\
whisper large                    & \textbf{79.58} / \textbf{83.33} / \textbf{82.50}        & 72.50 / \textbf{82.08} / 74.17   & \textbf{62.08} / 68.33 / \textbf{70.83} & \textbf{67.50} / 68.75 / 65.42        \\
whisper large v2                 & \textbf{85.42} / \textbf{84.58} / \textbf{80.42}        & \textbf{80.83} / 77.92 / 77.08  & \textbf{60.42} / 69.58 / \textbf{70.42} & \textbf{59.58} / \textbf{75.42} / 63.33         \\
whisper large v3                 & 77.08 / 79.17 / 80.00          & 77.08 / 78.33 / 75.83        & \textbf{60.42} / 70.42 / \textbf{70.83}    & \textbf{61.67} / \textbf{77.50} / 64.17   \\         
\bottomrule
\end{tabular}}
\vspace{-0.7cm}
\end{table}

\vspace{-0.5cm}
\section{Experiments and results}
\vspace{-0.2cm}
%怎么做的实验，实验结果，主要说明ASR trans某些大于Manual，以及分析ASR trans相比manual的cross有啥突出的特征
\subsection{Experimental setup}
\vspace{-0.2cm}
For fine-tuning the ASR models, we employed the AdamW optimizer with a learning rate of \( 1 \times 10^{-5} \) and weight decay of \( 5 \times 10^{-3} \). The models were trained for 20 epochs with a batch size of 8, and performance was evaluated using WER. For AD detection, we used AdamW with a learning rate of \( 4 \times 10^{-4} \) and weight decay of \( 1 \times 10^{-5} \). The training ran for 50 epochs with a batch size of 16, using cross-entropy loss. To ensure robustness, we conducted 10 trials with different random seeds and evaluated performance based on mean accuracy. All experiments were performed on NVIDIA RTX 4090 or A100 GPUs.
\vspace{-0.6cm}
\subsection{ASR models performance}
\vspace{-0.2cm}
% Table~\ref{tab:ASR models performance} presents the mean WER of original and fine-tuned 18 ASR models on the ADReSS dataset. It can be observed that all ASR models achieved significant performance improvements after fine-tuning. The fine-tuned \textit{whisper large v3} attained the lowest WER. Regardless of the WER values, the 36 ASR transcripts from the ADReSS will be utilized in AD detection research.
Table~\ref{tab:ASR models performance} shows the mean WER of original and fine-tuned 18 ASR models on the ADReSS dataset. It can be observed that all ASR models achieved significant performance improvements after fine-tuning, with the fine-tuned \textit{whisper large v3} achieving the lowest WER. Regardless of the WER values, the 36 ASR transcripts from ADReSS will be used in AD detection research.
\vspace{-0.6cm}
\subsection{AD detection results}
\vspace{-0.2cm}
% \subsubsection{Text-based results}
%表3展示了transcripts和speech上的AD检测结果. 从这个表中，我们能观察到: (a) 在每一种AD检测方法中, 都存在某些ASR转录(或者合成的语音)在准确率方面优于手动转录(或者合成的语音), 表明ASR错误可能会为AD检测提供有价值的cues，我们猜测这些ASR在AD和HC组中通过不对称的错误引入了系统性的偏差，这些偏差提供了有价值的cues，模型利用了这些cues来改进了AD检测准确率. (b) 在原始的语音和手动转录上，提出的基于corss-attention的方法相比基于self-attention的方法实现了更优或者可比的性能，同时又能提供可解释性，证明了该方法的优势与有效性. (c) 在每一种AD检测方法中，合成的的语音获得的AD检测准确率都小于原始的语音获得的AD检测准确率，表明CosyVoice2这个TTS模型合成的语音难以捕获AD患者原始发音中的全部病理特征. (d) 在每一种AD检测方法中，手动转录获得的准确率均优于原始的speech获得的准确率，表明仅利用语言学features优于仅利用声学特征,这与之前的发现一致\cite{syed2020automated,li2021comparative}. (e) 在基于self-attention的方法中，pre-trained embeddings获得的准确率几乎都大于knowledge-based feature获得的准确率，表明高纬度的embedding能捕获 more comprehensive, rich,and hierarchical representations相比knowledge-based feature.
Table~\ref{tab:AD detection results} presents the AD detection results on transcripts and speech. Key observations include: (a) In each detection method, certain ASR transcripts (ASR-synthesized speech) outperform manual transcripts (manual-synthesized speech) in accuracy, suggesting ASR errors may offer valuable cues for AD detection. We hypothesize that these errors introduce asymmetric biases between AD and HC groups, which provide useful cues that the model may exploit to improve accuracy. (b) The proposed cross-attention-based method outperforms or achieves comparable performance to the self-attention-based method on both original speech and manual transcripts, while also offering interpretability, highlighting its effectiveness and advantages. (c) Accuracy from synthesized speech is consistently lower than from original speech in all methods, indicating that the CosyVoice2 TTS model fails to fully capture pathological features in AD patients’ speech. (d) Accuracy from manual transcripts exceeds that from original speech in all methods, suggesting linguistic features are more effective than acoustic features alone in AD detection. (e) In the self-attention-based method, pre-trained embeddings generally outperform knowledge-based features, showing that pre-trained embeddings capture richer, more comprehensive representations.
\vspace{-0.3cm}
\subsection{Cross-attention-based interpretability analysis}
\vspace{-0.2cm}
%为了探究这些ASR errors提供的有价值的线索到底是什么，我们利用cross-attention-based可解性框架分别从转录和合成的语音的角度进行分析。具体来说，对于转录，我们选取了使用corss-attention-based方法在ASR转录上表现最好的单个seed得到的模型, 然后将ADReSS测试集输入这个模型，将正确预测的样本得到attention scores (也就是Section~ref{sec:Cross-attention-based interpretability model}中的 matrix A)进行平均，如Figure~ref{fig:asr manual text comparison}中的子图(a) 所示, 该子图横轴0-34分别对应于Section~ref{sec:Knowledge-based features}中按顺序介绍的每一个特征，该图揭示了分类模型对不同的knowledge-based 特征关注程度的大小以及pre-trained embeddings和knowledge-based 特征之间的整体关系. 类似的做法应用于Manual transcript, 得到 Figure~ref{fig:asr manual text comparison}中的子图(b). 我们还将上述两个子图在embedding维度进行求和再归一化得到子图(c)，以更好地形成对比来揭示ASR转录相比手动转录哪些Knowledge-based features的attention更高，这些特征可能就作为了有价值的线索来改进AD检测。
We explored the valuable cues provided by ASR errors from two perspectives: transcript and synthesized speech. Specifically, we compared and analyzed the attention scores generated by best-performing classification models on ASR and manual transcripts (or ASR- and manual-synthesized speech).

\begin{figure}[t!]
% \vspace{-0.3cm}
    \centering
    % 第一行子图
    \begin{subfigure}[htb]{0.23\textwidth}
        \centering
        \includegraphics[width=\textwidth]{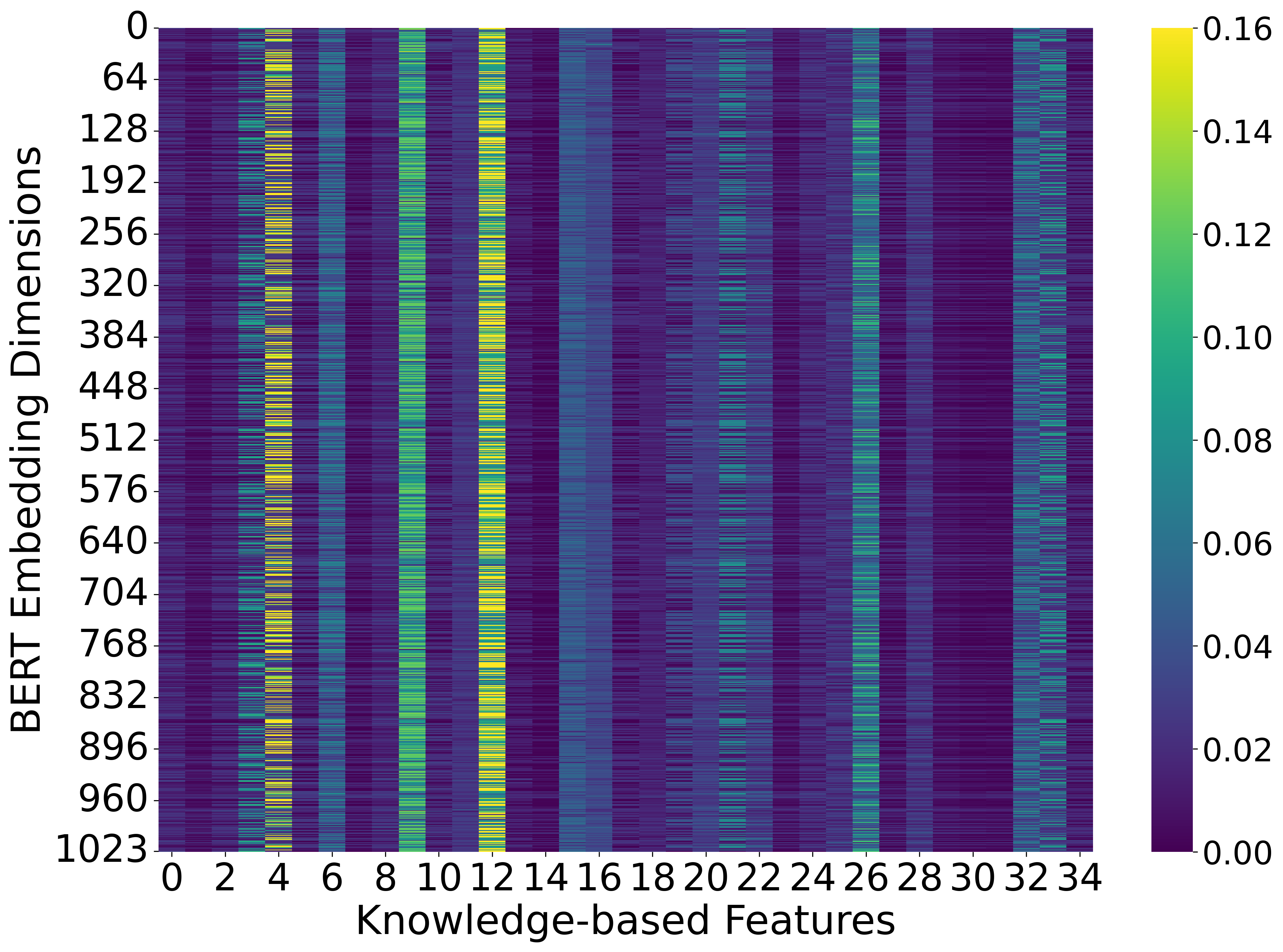}
        \caption{ASR transcripts attention.}
        \label{fig:asr text}
    \end{subfigure}
    \hspace{0.0001\textwidth}  % 调整水平间距
    \begin{subfigure}[htb]{0.23\textwidth}
        \centering
        \includegraphics[width=\textwidth]{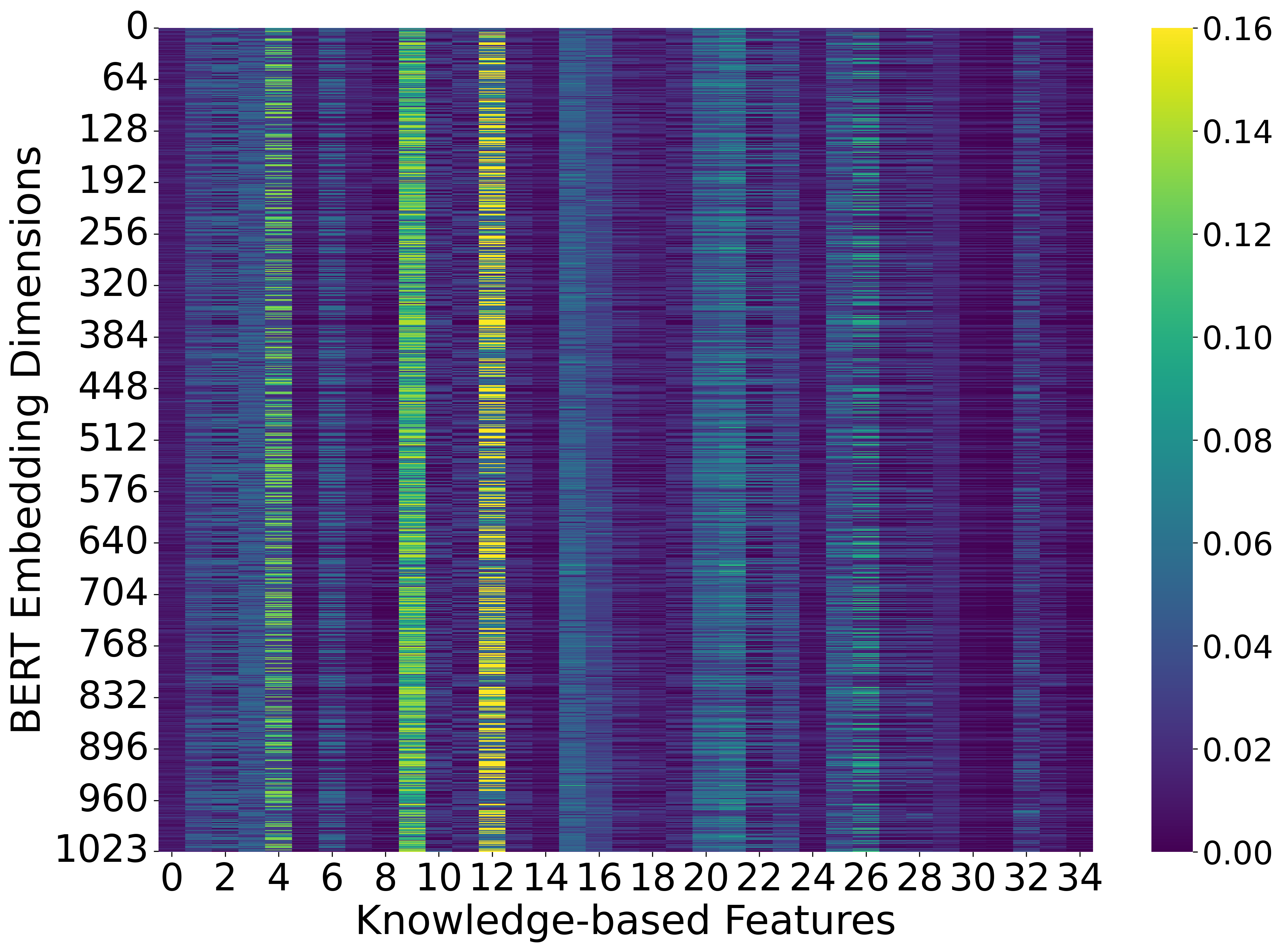}
        \caption{Manual transcripts attention.}
        \label{fig:manual text}
    \end{subfigure}
    
    \vspace{0.5em}  % 调整垂直间距
    
    % 第二行子图
    \begin{subfigure}[htb]{0.4\textwidth}
    \vspace{-0.3cm}
        \centering
        \includegraphics[width=\textwidth]{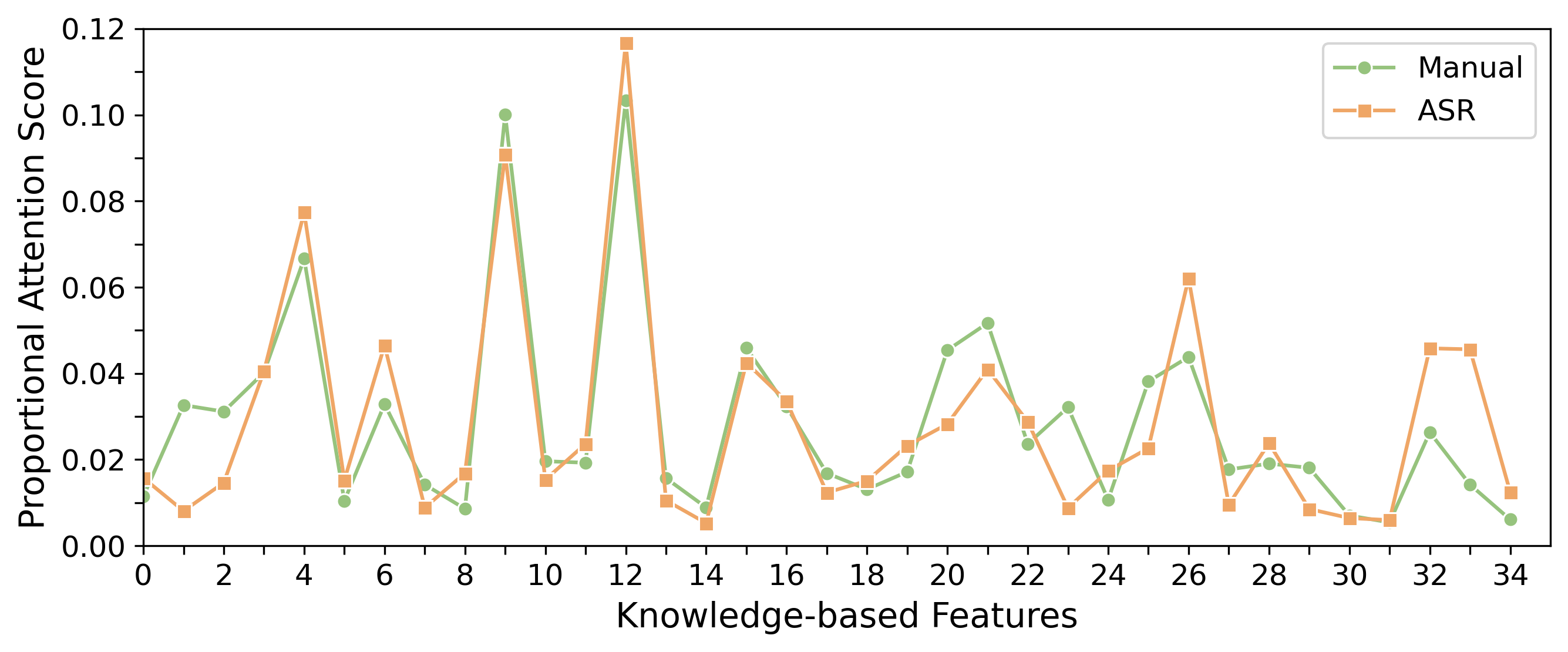}  % 新图片路径
        \caption{Attention comparison of ASR and manual transcripts.}
        \label{fig:asr manual text}
        \vspace{-0.3cm}
    \end{subfigure}
    %\hspace{0.0001\textwidth}  % 保持水平间距一致
    % \begin{subfigure}[htb]{0.23\textwidth}
    %     \centering
    %     \includegraphics[width=\textwidth]{manual_text/attn_informed_average_seed_52.png}  % 新图片路径
    %     \caption{Manual 2}
    %     \label{fig:manual text 2}
    % \end{subfigure}
    
    \caption{Attention analysis of transcripts.}
    \label{fig:asr manual text comparison}
    \vspace{-0.9cm}
\end{figure}

\textbf{Transcript Perspective}: We first selected the cross-attention-based model with the best-performing random seed on ASR transcripts (from fine-tuned \textit{whisper large}, with the best detection accuracy of 87.5\%), then input the corresponding ADReSS test set into this model. For each correctly predicted sample, we obtained an attention score matrix (i.e., matrix \(A\) in Section~\ref{sec:Cross-attention-based interpretability model}), then averaged all matrices, as shown in Figure~\ref{fig:asr text}. The x-axis (0-34)  corresponded to each of the knowledge-based text features introduced sequentially in Section~\ref{sec:Knowledge-based features}. This figure revealed the classification model's attention to different knowledge-based features and  the relationship between pre-trained embeddings and knowledge-based features. A similar approach was applied to manual transcripts (the best detection accuracy was 83.33\%), resulting in Figure~\ref{fig:manual text}. We summed and normalized the two figures across the embedding dimension to obtain Figure~\ref{fig:asr manual text}, providing a clearer contrast that highlighted which knowledge-based features received higher attention in ASR transcripts compared to manual transcripts. From Figure~\ref{fig:asr manual text}, the primary features with higher attention in ASR transcripts include syllable count, lexicon count, average sentence length, filler pauses ratio, repetition words ratio, FRE, GFI, DCRS, concreteness, imagability, meaningfulness, and age of acquisition rating. This suggests that these features may show greater differences between the AD and HC groups in ASR transcripts, compared to manual transcripts, making them more discriminative and valuable cues for improving AD detection. 
Our additional statistical experiments support the hypothesis mentioned above: compared to manual transcripts, the AD group in ASR transcripts exhibits lower syllable count, lexicon count, and average sentence length, along with higher filler pauses and repetition words ratios. These ASR errors reflect the decline in language abilities of AD patients (e.g., more indistinct pronunciation and disfluencies) and amplify key linguistic features, indicating that ASR could serve not only as a transcription tool but also as an auxiliary tool for AD detection. Moreover, designing ASR models tailored for AD-related speech could enhance this effect.

\begin{figure}[tb!]
% \vspace{-0.3cm}
    \centering
    % 第一行子图
    \begin{subfigure}[htb]{0.23\textwidth}
        \centering
        \includegraphics[width=\textwidth]{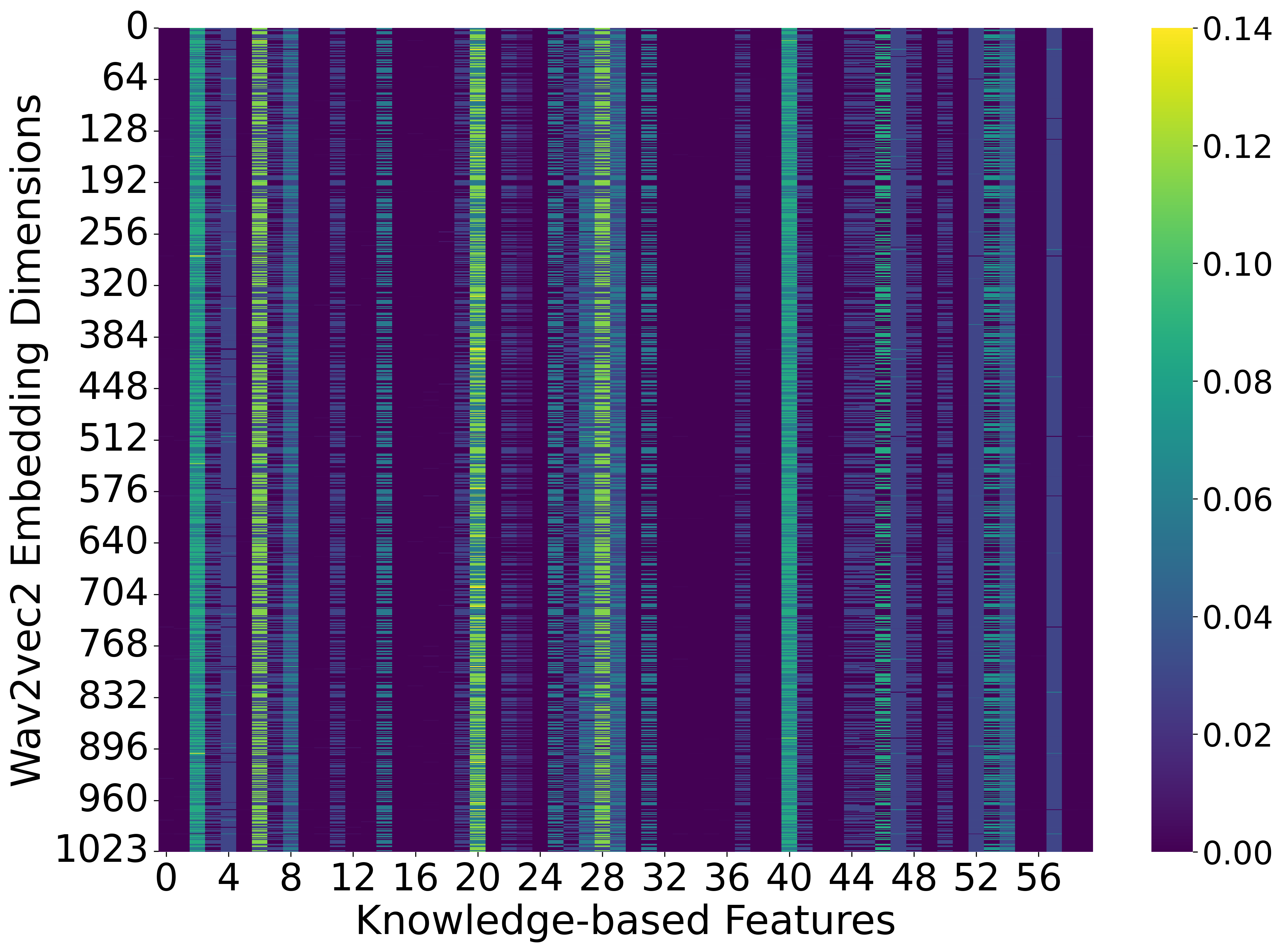}
        \caption{Attention of speech synthesized from ASR transcripts.}
        \label{fig:asr audio}
    \end{subfigure}
    \hspace{0.0001\textwidth}  % 调整水平间距
    \begin{subfigure}[htb]{0.23\textwidth}
        \centering
    \includegraphics[width=\textwidth]{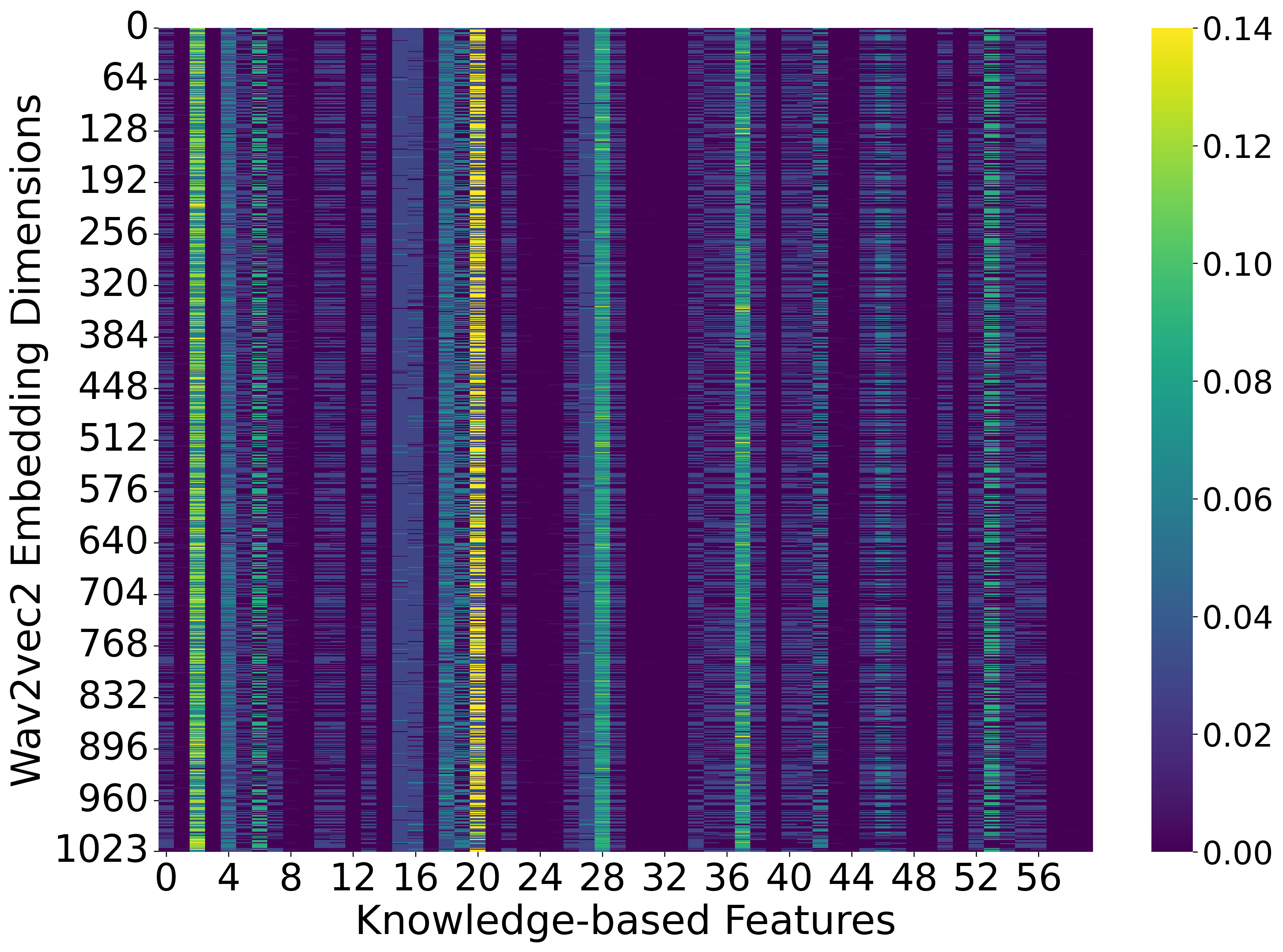} 
        \caption{Attention of speech synthesized from manual transcripts.}
        \label{fig:manual audio}
    \end{subfigure}
    
    \vspace{0.5em}  % 调整垂直间距
    
    % 第二行子图
    \begin{subfigure}[htb]{0.4\textwidth}
    \vspace{-0.3cm}
        \centering
        \includegraphics[width=\textwidth]{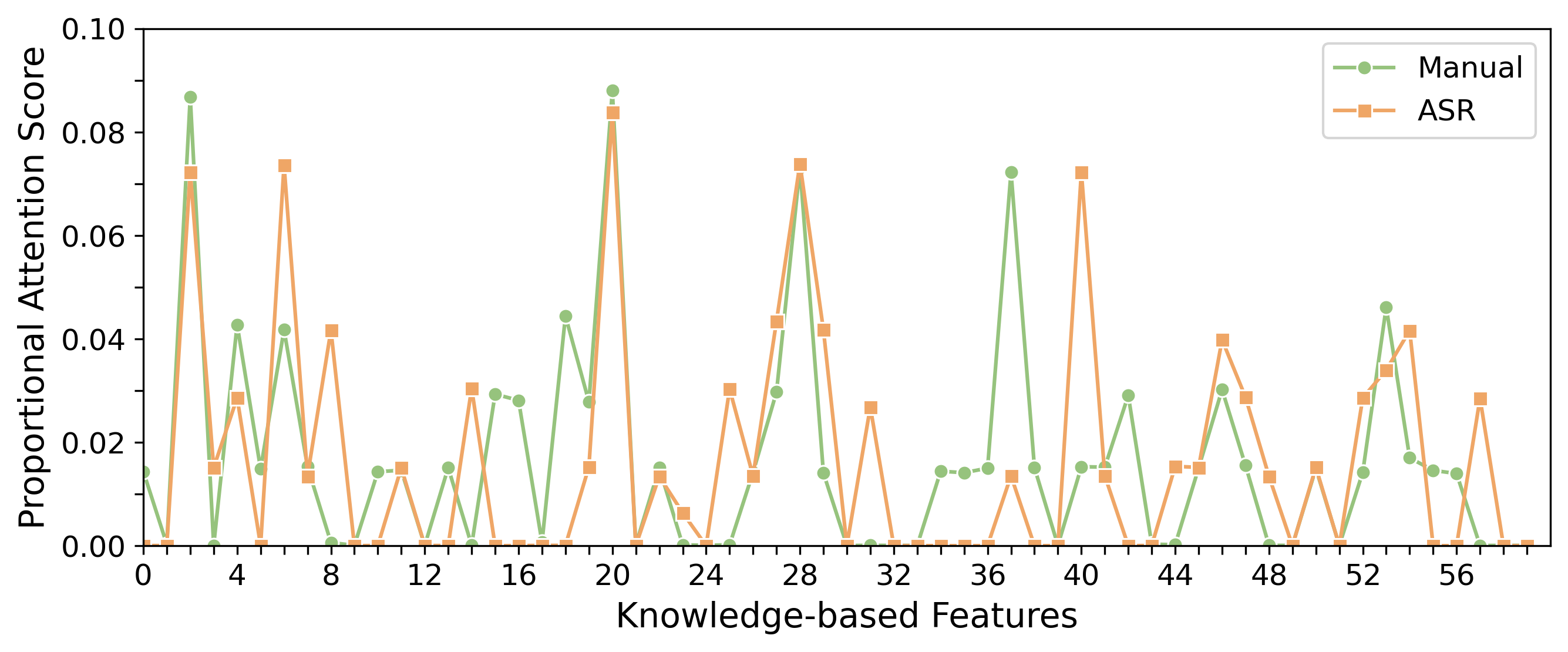}  % 新图片路径
        \caption{Attention comparison of speech synthesized from ASR and manual transcripts.}
        \label{fig:asr manual audio}
        \vspace{-0.3cm}
    \end{subfigure}
    % \hspace{0.0001\textwidth}  % 保持水平间距一致
    % \begin{subfigure}[htb]{0.23\textwidth}
    %     \centering
    %     \includegraphics[width=\textwidth]{manual_audio/attn_informed_average_seed_9.png}  % 新图片路径
    %     \caption{Manual 2}
    %     \label{fig:manual_audio_2}
    % \end{subfigure}
    
    \caption{Attention analysis of synthesized speech.}
    \label{fig:asr manual audio comparison}
    \vspace{-0.9cm}
\end{figure}

\textbf{Synthesized Speech Perspective}: We applied the same approach as in the transcript perspective, resulting in Figure~\ref{fig:asr manual audio comparison}. The x-axis (0–59) corresponded to the knowledge-based speech features introduced sequentially in Section~\ref{sec:Knowledge-based features}. The best-performing model for ASR-synthesized speech, based on the fine-tuned \textit{whisper small}, achieved 77.08\% detection accuracy, while the best detection accuracy for manual-synthesized speech was 72.92\%. From Figure~\ref{fig:asr manual audio}, the primary features with higher attention in ASR-synthesized speech compared to manual-synthesized speech include the mean, maximum, and skewness of pause duration; the maximum, minimum, and skewness of unvoiced duration; maximum of voiced duration; PVU; VVU; mean, maximum, minimum, and skewness of F0; std of F2; mean of NAQ and OQ variability; and the std of GCI variability and OQ average. Similar to the analysis from the transcript perspective, the features mentioned above may serve as valuable cues for AD detection. Since ASR errors can partially reflect the dysfluencies of AD patients, dysfluency-related features in ASR-synthesized speech may play a more significant role, which aligns with the features related to pause, voiced, and unvoiced duration listed above.
\vspace{-0.3cm}
\subsection{AD-relevant patterns within pre-trained embeddings}
\vspace{-0.2cm}
We employed attention scores to unveil the knowledge-based features encoded in the pre-trained embeddings from both manual transcripts and original speech (focusing on intrinsic characteristics rather than comparative analysis with ASR). For manual transcripts, Figures~\ref{fig:manual text} and~\ref{fig:asr manual text} (green line) show that the repetition words ratio (12), content density (9), and syllable count (4) exhibit higher attention or correlation with the BERT embedding. For original speech, a similar attention score heatmap and line plot (omitted due to space limitations) reveal that PVU, along with the maximum and minimum voiced duration, exhibit higher attention or correlation with the Wav2Vec2 embeddings. These findings highlight the significance of these features in AD detection. Additionally, we found that almost all knowledge-based features are not encoded into specific dimensions of the pre-trained embeddings, but instead emerge through complex interrelationships across nearly all dimensions, suggesting that almost all embedding dimensions contribute significantly.
%对于手动转录, 我们从Figure~\ref{fig:asr manual text comparison}中的子图(b)和子图(c)中的绿线可以看出, repetition words ratio (12), content density (9), syllable count (4)这三种特征展示了与BERT embedding更高的注意力或者相关性。对于original speech，我们得到了类似的基于attention score 热力图和折线图(由于文章篇幅限制所以没有展示), 从图中可以看出PVU, the maximum, minimum, and std of voiced duration展示了与Wav2Vec2 embedding更高的注意力或者相关性. 以上发现也表明了这些特征在AD检测中重要性. 我们还观察到，对于几乎每一个knowledge-based特征，它们并不是被pre-trained embeddings编码到特定的维度，而是通过几乎所有维度的embeddings之间的复杂的相互关系来产生的，也就是说几乎所有的SSL嵌入维度都起着重要的作用。
\vspace{-0.3cm}
\section{Conclusions}
\vspace{-0.3cm}
%在本文中，我们进行了一个关于使用各种WER的ASR转录和它们的合成的语音进行阿尔茨海默症检测的综合研究。AD检测结果表明ASR错误可能会为提高AD检测提供有价值的线索。我们证实这些errors通过在AD和HC组之间引入不对称偏差来提供有用的线索，模型可以利用这些线索来提高AD检测的准确性。此外，我们提出了一种基于交叉注意力的可解释性模型，该模型不仅解释了这些有价值的线索是什么而且能实现更优或者可比的性能，我们还利用该模型揭示了预训练嵌入中的AD相关模式。这项研究为ASR系统在AD检测中的意外潜力提供了新的见解。
In this paper, we conducted a comprehensive study on AD detection using ASR transcripts with various WER and their corresponding synthesized speech. Our results indicated that ASR errors may offer valuable cues for improving AD detection, which we attributed to the asymmetric biases introduced by these errors between the AD and HC groups.
% . We believed that these cues arose from the asymmetric biases introduced by ASR errors between the AD and HC groups. 
Additionally, we proposed a cross-attention-based interpretability model that not only identified valuable cues from both transcript and speech perspectives but also achieved superior or comparable performance to the baseline model. We further leveraged this model to unveil AD-related patterns within pre-trained embeddings. This study provided new insights into the potential of ASR models in AD detection. Our future work will focus on designing ASR models tailored for AD speech to maximize this potential.
\section{Acknowledgements}
This work was partially supported by the Anhui Province Major Science and Technology Research Project (Grant No. S2023Z20004), by the National Social Science Foundation of China (Grant No. 23AYY012), and by the Supercomputing Center of the University of Science and Technology of China.
\bibliographystyle{IEEEtran}
\bibliography{mybib} % Ensure no conflicting biblatex commands are present in the document.
\end{document}